\documentclass[conference]{IEEEtran}
\IEEEoverridecommandlockouts
\usepackage{cite}
\usepackage{amsmath,amssymb,amsfonts}
\usepackage{algorithmic}
\usepackage{graphicx}
\usepackage{textcomp}
\usepackage{xcolor}
\usepackage{textpos}
\usepackage{booktabs}
\usepackage{multirow}
\usepackage{subfig}
\usepackage{url}
\usepackage{mathabx}
\usepackage{graphbox} 
\usepackage{array} 
\usepackage{arydshln} 
\usepackage{hyperref}
\usepackage{tikz}
\usepackage{pgf}
\usepackage[linesnumbered,ruled,vlined]{algorithm2e} 

\graphicspath{{./img/}}

\def\BibTeX{{\rm B\kern-.05em{\sc i\kern-.025em b}\kern-.08em
    T\kern-.1667em\lower.7ex\hbox{E}\kern-.125emX}}
\begin{document}

\title{
Generating 3D Terrain with 2D Cellular Automata
\thanks{Short paper. This work was partially funded by: Fundação para a Ciência e a Tecnologia (FCT) under grants CEECINST/00002/2021/CP2788/CT0001, UIDB/00066/2020, UIDB/04111/2020, and UIDB/05380/2020; and, Instituto Lusófono de Investigação e Desenvolvimento (ILIND) under Project\linebreak COFAC/ILIND/COPELABS/1/2024.}}

\author{\IEEEauthorblockN{
Nuno Fachada\IEEEauthorrefmark{1}\IEEEauthorrefmark{2}, 
António R. Rodrigues\IEEEauthorrefmark{3}, 
Diogo de Andrade\IEEEauthorrefmark{3}, 
and
Phil Lopes\IEEEauthorrefmark{3}} 
\IEEEauthorblockA{\IEEEauthorrefmark{1}Lusófona University, ECATI, 
Campo Grande, 376, 1749-024 Lisboa, Portugal\\
Email: nuno.fachada@ulusofona.pt}
\IEEEauthorblockA{\IEEEauthorrefmark{2}
Center of Technology and Systems (UNINOVA-CTS) and Associated Lab of Intelligent Systems (LASI),\\
2829-516 Caparica, Portugal}
\IEEEauthorblockA{\IEEEauthorrefmark{3}Lusófona University, HEI‐Lab: Digital Human‐Environment Interaction Labs, 
Campo Grande, 376, 1749-024 Lisboa, Portugal\\
Email: a22202884@alunos.ulht.pt, diogo.andrade@ulusofona.pt, phil.lopes@ulusofona.pt}
}

\maketitle

\begin{textblock*}{200mm}(-1cm,-7.5cm)
  \noindent \footnotesize The peer-reviewed version of this paper is
  published in IEEE Xplore at
  \href{https://doi.org/10.1109/CoG64752.2025.11114361}{\texttt{https://doi.org/10.1109/CoG64752.2025.11114361}}.
  This version is typeset by the authors and differs only in pagination and
  typographical detail.
\end{textblock*}

\begin{abstract}
This paper explores the use of 2D cellular automata (CA) to generate 3D terrains through a simple additive approach. Experimenting with multiple CA transition rules produced aesthetically interesting, navigable landscapes, suggesting applicability for terrain generation in games.

\end{abstract}


\definecolor{gray0}{rgb}{0.9,0.9,0.9}
\definecolor{gray1}{rgb}{0.85,0.85,0.85}
\definecolor{gray15}{rgb}{0.785,0.785,0.785}
\definecolor{gray2}{rgb}{0.7,0.7,0.7}
\definecolor{gray3}{rgb}{0.55,0.55,0.55}
\definecolor{gray4}{rgb}{0.4,0.4,0.4}

\newcommand{\drawGrid}{
    \foreach \y in {0,...,5} {
        \foreach \x in {0,...,5} {
            \pgfmathparse{\castate[\y][\x]}
            \ifnum\pgfmathresult=1
                \fill[gray15] (\x, 5-\y) rectangle ++(1,1);
            \fi
            \draw[gray0] (\x, 5-\y) rectangle ++(1,1);
        }
    }
}

\newcommand{\drawFinalGrid}{
    \foreach \y in {0,...,5} {
        \foreach \x in {0,...,5} {
            \pgfmathparse{\castate[\y][\x]}
            \ifnum\pgfmathresult=1
                \fill[gray1] (\x, 5-\y) rectangle ++(1,1);
            \fi
            \ifnum\pgfmathresult=2
                \fill[gray2] (\x, 5-\y) rectangle ++(1,1);
            \fi
            \ifnum\pgfmathresult=3
                \fill[gray3] (\x, 5-\y) rectangle ++(1,1);
            \fi
            \ifnum\pgfmathresult=4
                \fill[gray4] (\x, 5-\y) rectangle ++(1,1);
            \fi
            \draw[gray0] (\x, 5-\y) rectangle ++(1,1);
        }
    }
}

\section{Introduction}
\label{sec:intro}

Procedural generation of 3D landscapes and terrains is an important aspect of game development, allowing for unique and expansive environments while fostering replayability. Notwithstanding issues with unpredictability and possibly incoherent gameplay experiences that may require extensive testing, procedural terrain generation also has the potential to optimize resources and reduce storage costs.

In this paper we present an initial exploration on the use of 2D cellular automata (CA) for the purpose of generating 3D terrains. The proposed method is deceptively simple yet 
novel and able to produce aesthetically interesting landscapes.

The paper is organized as follows. In Section~\ref{sec:background}, we present some background of previous work in this field. In Section~\ref{sec:methods}, the proposed novel CA terrain generation method is described. Results, discussion, and limitations follow in Section~\ref{sec:results}. Finally, in Section~\ref{sec:conclusions}, we draw some conclusions and outline future work.

\section{Background}
\label{sec:background}

Heightmap-based methods are commonly used for terrain generation, where greyscale textures represent elevation data. These heightmaps can be manually crafted or procedurally generated using various noise and fractal functions~\cite{shaker2016fractals}. 
On the other hand, CA are discrete, abstract computational systems characterized by a regular grid of cells, each in one of a finite number of states, which evolve in discrete time steps according to a set of rules based on the states of neighboring cells \cite{toffoli1987cellular}. They have been previously used for terrain generation in games, namely caves in 2D maps \cite{johnson2010cellular}, levels for real-time strategy games \cite{ziegler2020generating}, or natural-looking generic 2D game maps \cite{wu2021procedural}, among others.

\section{Methods}
\label{sec:methods}

The proposed method is simple. A binary grid is randomly initialized with 50\% probability per cell, representing the initial 2D CA state. The CA evolves by a specified transition rule~\cite{toffoli1987cellular} over $i_{\max}$ iterations. States are summed into a heightmap with values between 0 and $i_{\max}$, as shown in Fig.~\ref{fig:stackca}. The random initial state can often be discarded to reduce irregularities. Finally, the heightmap is rescaled to $[0, h_{\max}]$ using min-max normalization. This process is summarized in Algorithm~\ref{alg:hmgen}.

\begin{figure*}[t]
    \setlength{\tabcolsep}{0.05cm}
    \renewcommand{\arraystretch}{1.5}
    \begin{center}
    \begin{tabular}{ccccccccccc}

    \raisebox{-.45\height}{
    \begin{tikzpicture}[scale=0.35]
        \def\castate{{{1,0,1,0,1,1},{0,0,0,0,0,0},{1,0,1,0,0,1},{1,1,1,0,0,1},{1,1,1,0,1,0},{0,0,1,0,1,1}}}
        \drawGrid
    \end{tikzpicture}
    }
    & $\rightarrow$ &
    \raisebox{-.45\height}{
    \begin{tikzpicture}[scale=0.35]
        \def\castate{{{0,1,0,0,0,0},{0,0,0,1,1,1},{0,1,0,0,0,0},{1,1,0,0,1,1},{1,1,0,1,0,1},{1,1,0,1,0,1}}}
        \drawGrid
    \end{tikzpicture}
    }
    & $+$ &
    \raisebox{-.45\height}{
    \begin{tikzpicture}[scale=0.35]
        \def\castate{{{1,0,0,1,1,0},{1,0,1,0,0,0},{1,0,1,1,1,1},{1,0,0,0,0,0},{1,1,1,0,1,1},{1,0,1,0,0,0}}}
        \drawGrid
    \end{tikzpicture}
    }
    & $+$ &
    \raisebox{-.45\height}{
    \begin{tikzpicture}[scale=0.35]
        \def\castate{{{0,1,1,0,0,0},{0,1,0,1,1,1},{0,1,0,0,0,0},{1,1,0,1,1,1},{0,1,0,1,0,0},{0,1,0,1,0,1}}}
        \drawGrid
    \end{tikzpicture}
    }
    & $+$ &
    \raisebox{-.45\height}{
    \begin{tikzpicture}[scale=0.35]
        \def\castate{{{1,0,1,0,1,1},{0,0,1,0,0,0},{1,0,1,0,1,1},{0,0,1,0,0,0},{1,0,1,0,1,0},{0,0,1,0,1,0}}}
        \drawGrid
    \end{tikzpicture}
    }
    & $=$ &
    \raisebox{-.45\height}{
    \begin{tikzpicture}[scale=0.35]
        \def\castate{{{2,1,2,0,2,1},{0,0,1,1,1,1},{2,1,2,0,1,2},{3,2,1,0,1,2},{3,3,2,1,2,1},{1,1,2,1,1,2}}}
        \drawFinalGrid
    \end{tikzpicture}
    }\\
    $i=0$ & & $i=1$ & & $i=2$ & & $i=3$ & & $i=4$ & & Heightmap\\
  \end{tabular}
  \end{center}
  \caption{Creating a heightmap by adding sequential CA iterations. In this example, iterations 1 to 4 of a toroidal CA evolved with the Diamoeba rule (see Table~\ref{tab:rules}) are added, generating the heightmap on the right. The initial state ($i=0$), usually composed of random noise, can be discarded to avoid irregularities in the final heightmap, as done in this example.}
  \label{fig:stackca}
\end{figure*}

\begin{algorithm}
\DontPrintSemicolon 
\SetAlgoLined 
\BlankLine 

Initialize CA grid (e.g., with random noise)\;
Initialize heightmap to zeros\;

\For{$i = 1$ \KwTo $i_{\max}$}{
  Apply selected CA transition rule\;
  Add current CA grid state to heightmap\;
}

Rescale heightmap to $[0, h_{\max}]$ via min-max\;

\caption{Heightmap generation with 2D CA.} 
\label{alg:hmgen}
\end{algorithm}

The CA transition rules tested here are described in Table~\ref{tab:rules}. The Majority rule---the first listed---is simple: a cell with $N$ or more live neighbors in a Moore neighborhood of radius $r$ survives; otherwise, it dies~\cite{toffoli1987cellular}. The Caves rule, well known in PCG research, was proposed by Johnson et al. to generate 2D cave levels in real time~\cite{johnson2010cellular}. It is a slightly tweaked majority rule, in which a cell is able to survive (i.e., considering it is already alive) by having one less live neighbor than what it requires for being born (i.e., if it is dead). The Diamoeba rule has a gap in the birth neighbor interval, forming large, irregular diamond shapes in 2D~\cite{gravner1998cellular}. All rules assume a Moore neighborhood, with adjustable radius $r$ for Majority and Caves, and $r=1$ for Diamoeba.

\begin{table}
\caption{CA transition rules tested in this work.} 
\label{tab:rules}

\begin{center}
\begin{tabular}{ll}
\toprule
Rule & Description \\
\midrule
Majority $rN$ & Cell survives if $n \geq N$, is born if $n \geq N$. \\ 
Caves $rN$ & Cell survives if $n \geq N-1$, is born if $n \geq N$. \\ 
Diamoeba & Cell survives if $n \geq 5$, is born if $n \in \{3,5,6,7,8\}$.  \\ 
\bottomrule
\multicolumn{2}{p{0.95\columnwidth}}{\scriptsize\textbf{Note}:  The Moore neighborhood radius is denoted by $r$ (set to 1 by default), while $n$ represents the number of live neighbors. In the descriptions, \textit{survive} means the cell will continue to live if already alive, while being \textit{born} indicates that the cell will become alive when it was previously dead.}

\end{tabular}
\end{center}

\end{table}

The experiments were carried out using heightmaps with a resolution of $129 \times 129$, corresponding to the CA dimensions. The various $h_{\max}$ values presented consist of a percentage of 129. For the purpose of counting neighbors, the CA grid is considered toroidal. Evaluation was performed subjectively, by analysing and discussing prominent generated terrain features, and objectively, using four established terrain metrics: 1) the roughness index, which quantifies local elevation variability; 
2) normalized Shannon entropy, which reflects the diversity and distribution of elevation values; 3) slope walkability percentage, defined as the proportion of terrain with slopes below a climbable threshold set to 30º; and, 4) the path coverage percentage, which represents the fraction of walkable cells that belong to the largest connected region. For the latter two metrics, the radius of the pathfinding agent was set to 0.01\% of the length of the side of the terrain. Together, these metrics offer a quantitative view of terrain smoothness, structural richness, and potential for in-game exploration~\cite{yannakakis2018artificial}.

An implementation of these methods is available in the
Game AI Prototypes package \cite{fachada2023active},
developed in the Unity game engine~\cite{unity3d}.

\setlength{\dashlinedash}{0.5pt}
\setlength{\dashlinegap}{1.5pt}
\setlength{\arrayrulewidth}{0.3pt}
\newcolumntype{R}[1]{>{\raggedleft\arraybackslash}p{#1}}

\newcommand{\metic}[5]{%
\resizebox{3.16cm}{!}{%
    \centering
    \begin{tabular}{:R{0.7cm}:R{0.7cm}:R{0.85cm}:R{0.85cm}:}
      $R$ & $E$ & $W$ & $W_0$ \\
      #1 & #3 & #4 & #5 \\
    \end{tabular}%
  }%
}

\begin{figure*}

\begin{center}
\begin{tabular}{ccccc}

\addlinespace
\multicolumn{5}{l}{\textbf{Majority. $r=2,N=13,h_{\max}=5\%$}}\\
\includegraphics[width=3.25cm]{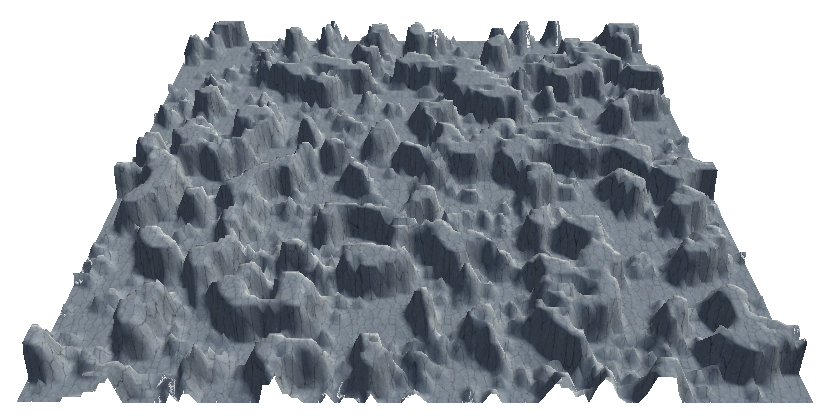}  &
\includegraphics[width=3.25cm]{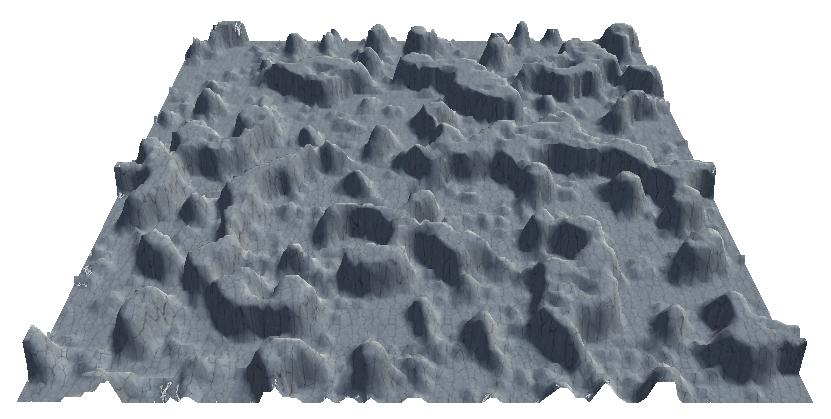}  &
\includegraphics[width=3.25cm]{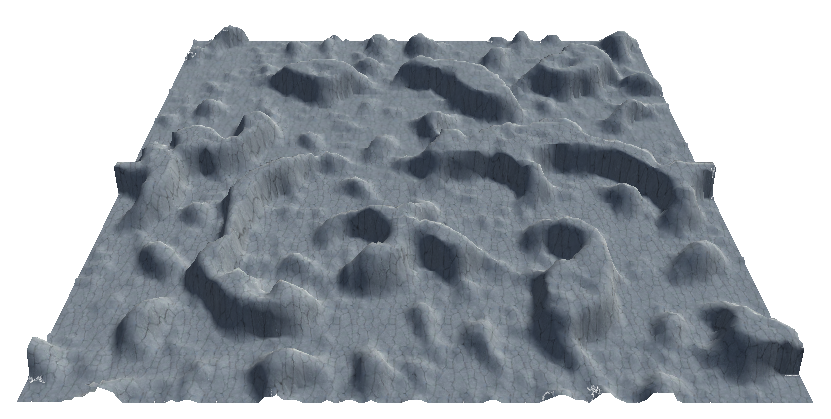}  &
\includegraphics[width=3.25cm]{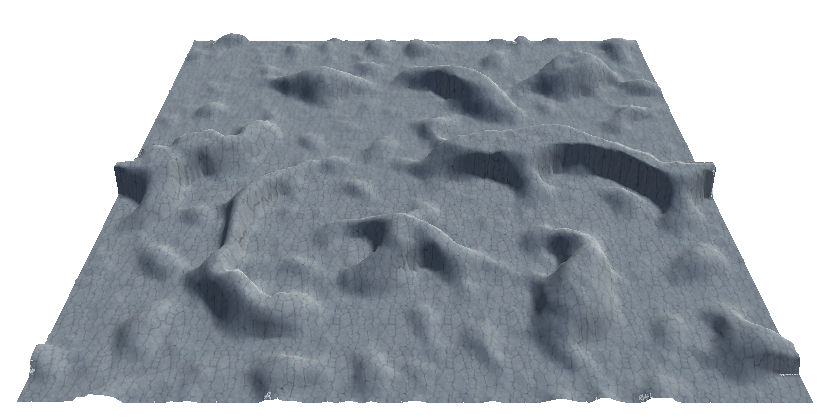}  &
\includegraphics[width=3.25cm]{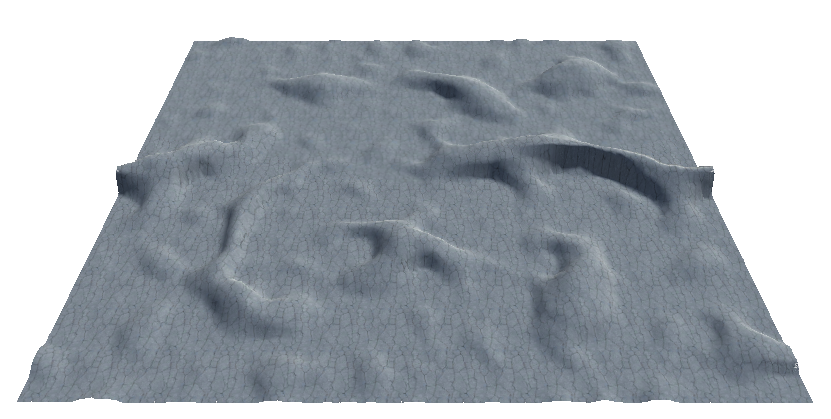}  \\
$i=3$ & $i=5$ & $i=10$ & $i=20$ & $i=50$ \\
\metic{ 0.011 }{ 1.635 }{ 0.211 }{ 29.4\% }{ 8.4\% }
&
\metic{ 0.009 }{ 1.650 }{ 0.248 }{ 37.4\% }{ 13.3\% }
&
\metic{ 0.006 }{ 1.670 }{ 0.301 }{ 67.2\% }{ 91.3\% }
&
\metic{ 0.004 }{ 1.685 }{ 0.345 }{ 77.5\% }{ 99.9\% }
&
\metic{ 0.003 }{ 1.691 }{ 0.366 }{ 90.5\% }{ 100.0\% }
\\

\addlinespace\addlinespace
\multicolumn{5}{l}{\textbf{Majority. $r=4,N=38,h_{\max}=10\%$}}\\
\includegraphics[width=3.25cm]{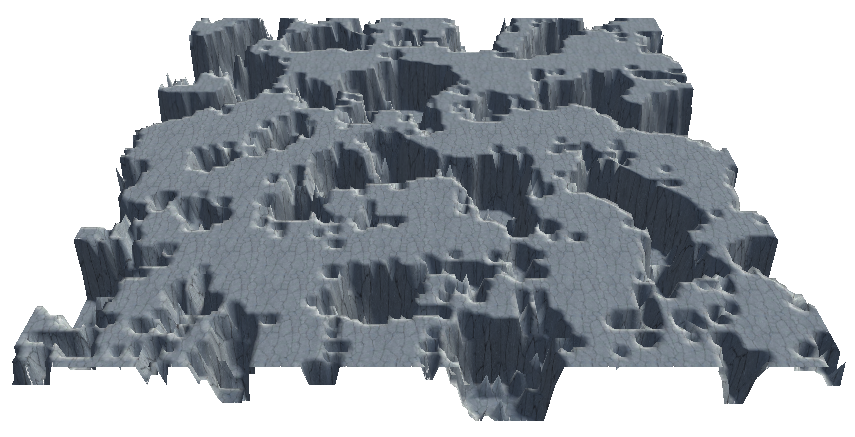}  &
\includegraphics[width=3.25cm]{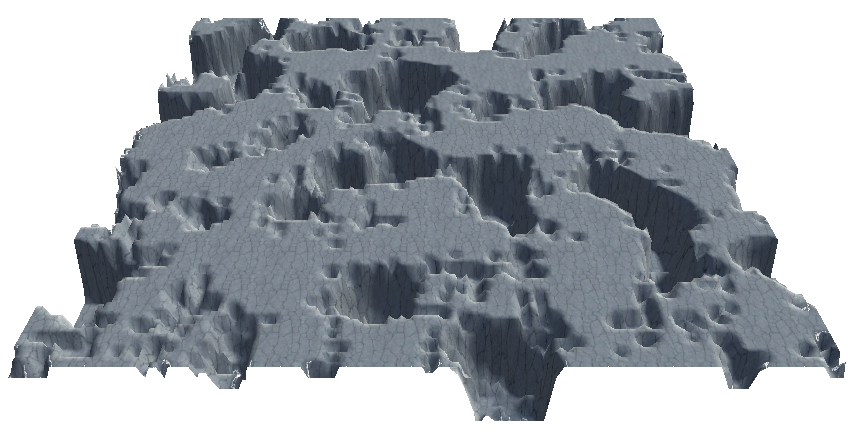}  &
\includegraphics[width=3.25cm]{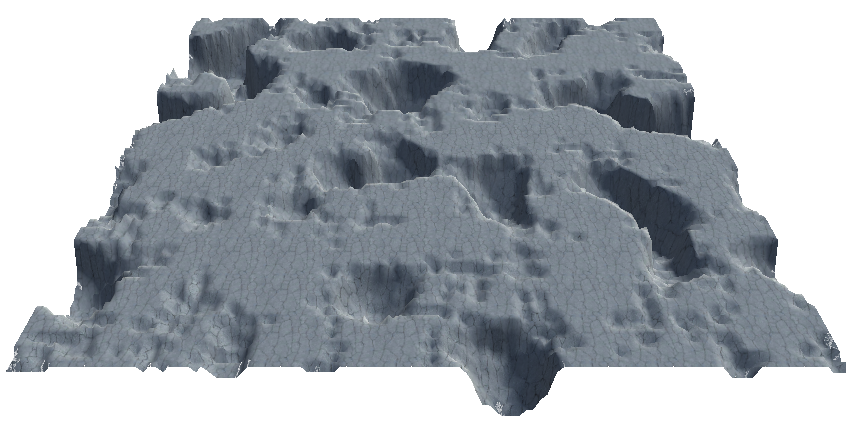}  &
\includegraphics[width=3.25cm]{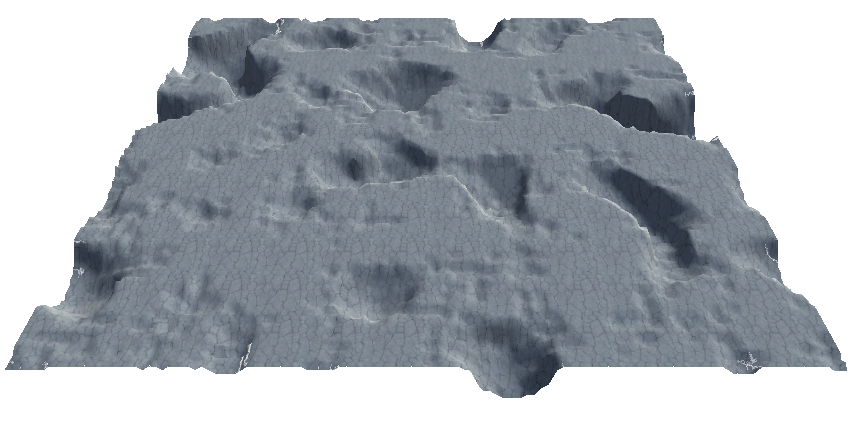} &
\includegraphics[width=3.25cm]{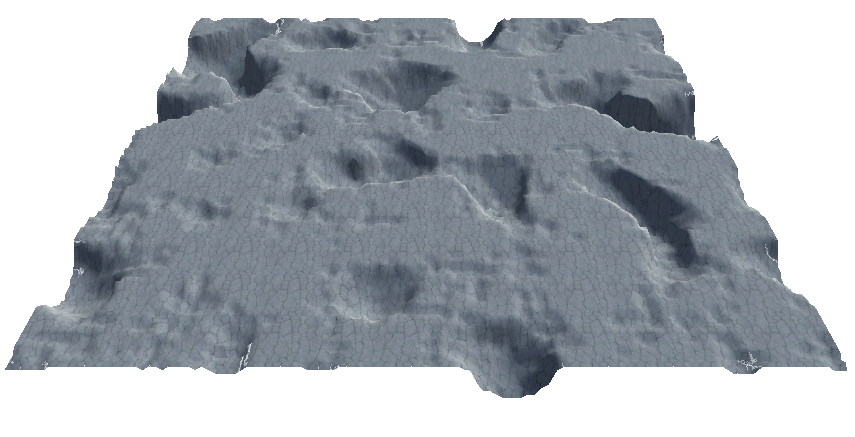} \\
$i=3$ & $i=5$ & $i=10$ & $i=20$ & $i=50$ \\
\metic{ 0.017 }{ 1.530 }{ 0.191 }{ 48.0\% }{ 76.1\% }
&
\metic{ 0.012 }{ 1.559 }{ 0.222 }{ 44.9\% }{ 81.4\% }
&
\metic{ 0.007 }{ 1.591 }{ 0.260 }{ 48.3\% }{ 85.0\% }
&
\metic{ 0.005 }{ 1.599 }{ 0.270 }{ 78.4\% }{ 97.2\% }
&
\metic{ 0.005 }{ 1.599 }{ 0.270 }{ 78.4\% }{ 97.2\% }
\\

\addlinespace\addlinespace
\multicolumn{5}{l}{\textbf{Caves. $r=2,N=13,h_{\max}=4\%$}}\\
\includegraphics[width=3.25cm]{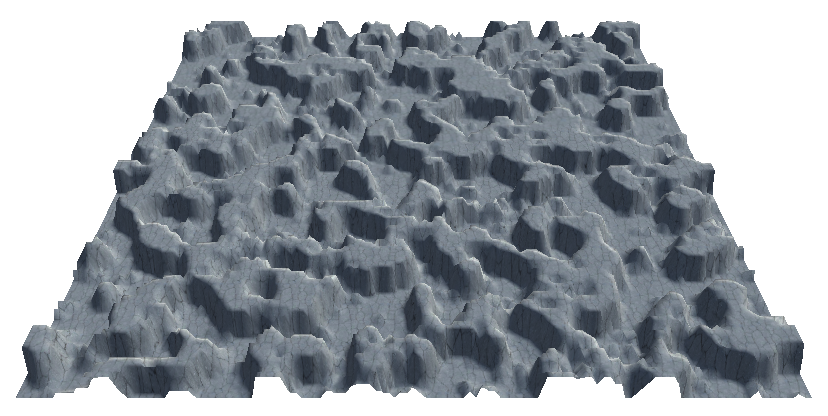}  &
\includegraphics[width=3.25cm]{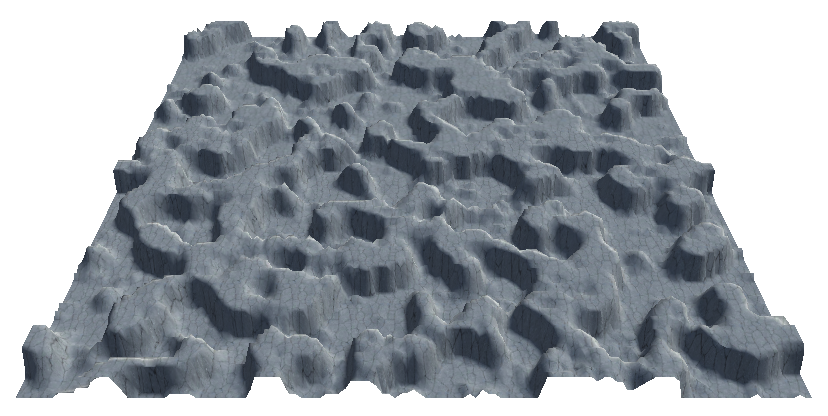}  &
\includegraphics[width=3.25cm]{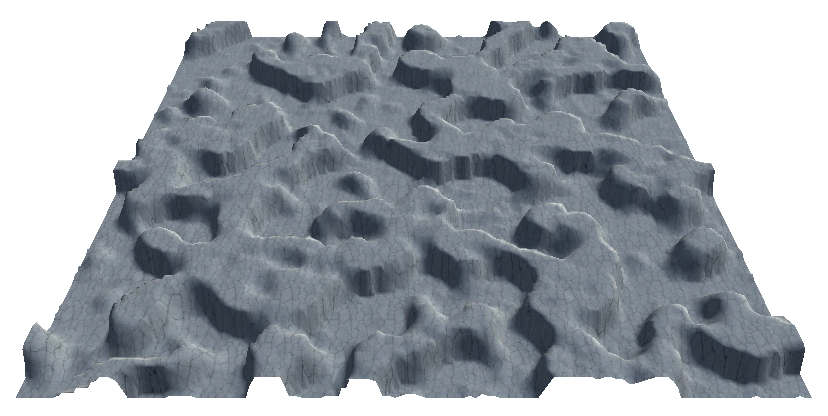}  &
\includegraphics[width=3.25cm]{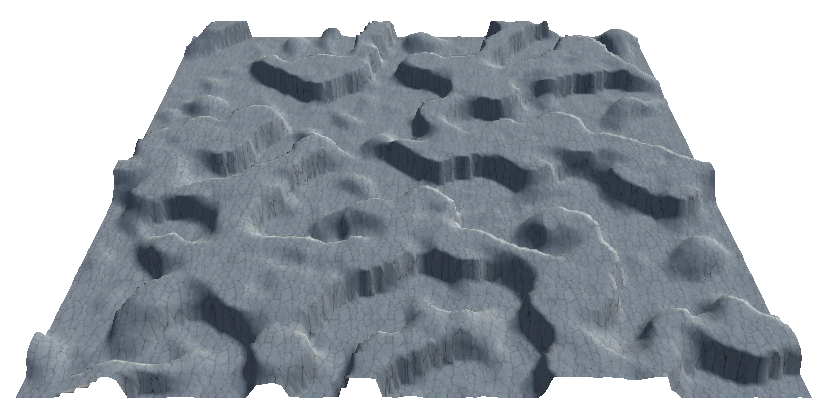} &
\includegraphics[width=3.25cm]{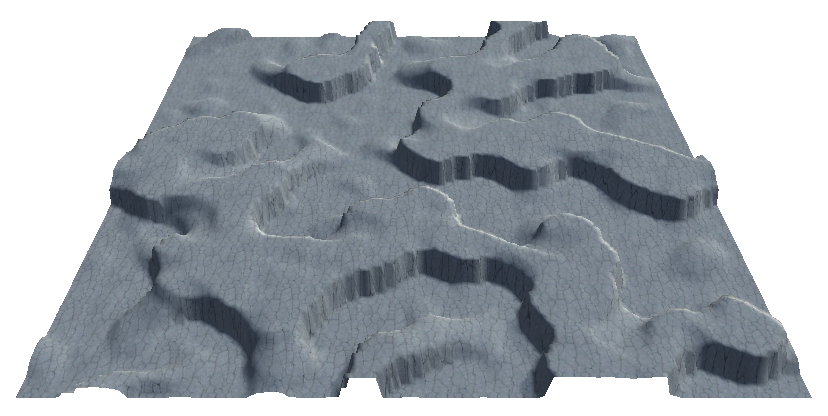} \\
$i=3$ & $i=5$ & $i=10$ & $i=20$ & $i=50$ \\
\metic{ 0.010 }{ 1.621 }{ 0.214 }{ 32.0\% }{ 5.0\% }
&
\metic{ 0.009 }{ 1.637 }{ 0.253 }{ 39.2\% }{ 5.4\% }
&
\metic{ 0.007 }{ 1.654 }{ 0.309 }{ 60.3\% }{ 34.4\% }
&
\metic{ 0.007 }{ 1.669 }{ 0.367 }{ 73.2\% }{ 94.0\% }
&
\metic{ 0.007 }{ 1.680 }{ 0.434 }{ 80.0\% }{ 84.8\% }
\\

\addlinespace\addlinespace
\multicolumn{5}{l}{\textbf{Diamoeba. $r=1,h_{\max}=5\%$}}\\
\includegraphics[width=3.25cm]{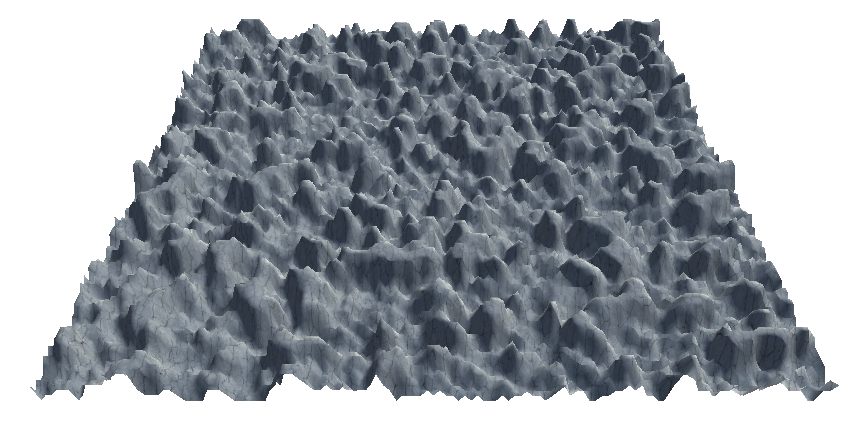}  &
\includegraphics[width=3.25cm]{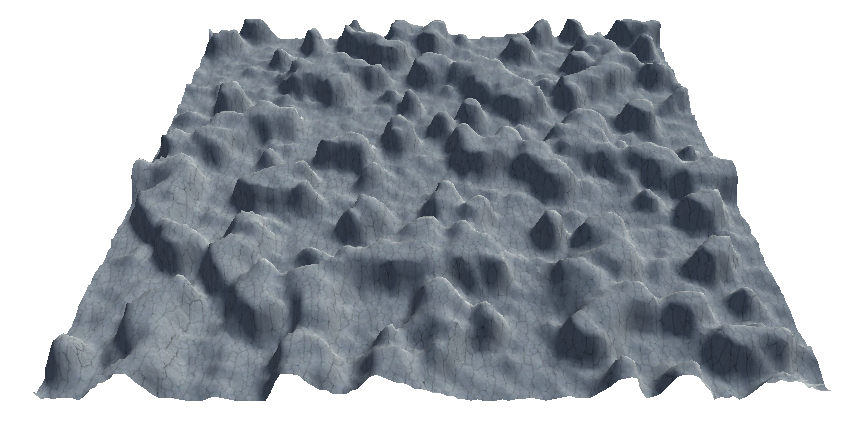}  &
\includegraphics[width=3.25cm]{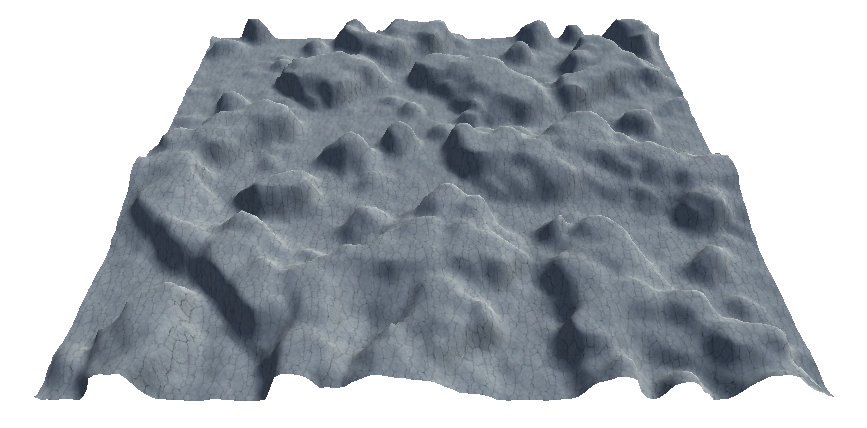}  &
\includegraphics[width=3.25cm]{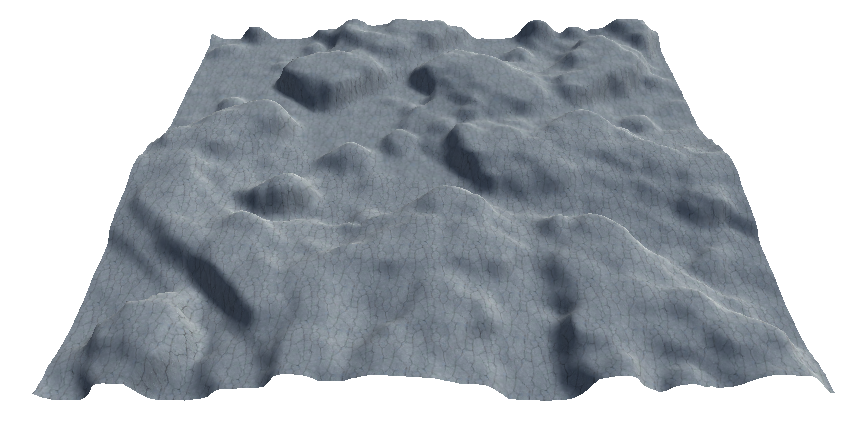}  &
\includegraphics[width=3.25cm]{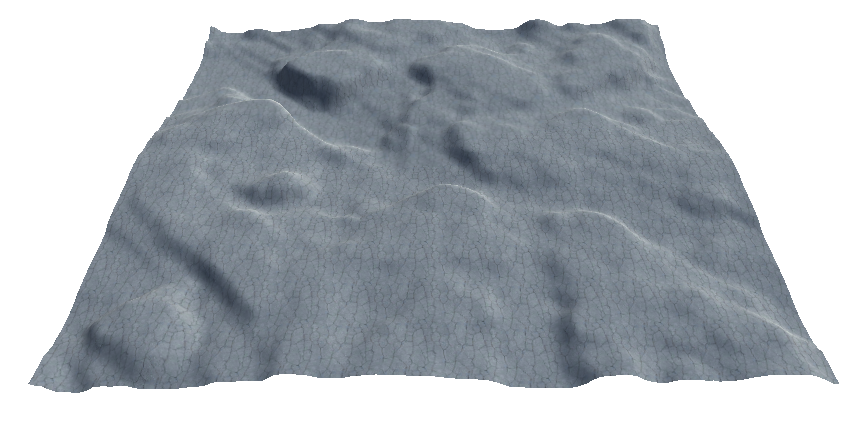}  \\
$i=5$ & $i=20$ & $i=50$ & $i=100$ & $i=200$ \\
\metic{ 0.009 }{ 1.784 }{ 0.309 }{ 5.2\% }{ 2.8\% }
&
\metic{ 0.006 }{ 1.794 }{ 0.500 }{ 45.5\% }{ 48.1\% }
&
\metic{ 0.004 }{ 1.799 }{ 0.641 }{ 66.2\% }{ 95.5\% }
&
\metic{ 0.003 }{ 1.803 }{ 0.777 }{ 81.2\% }{ 99.9\% }
&
\metic{ 0.002 }{ 1.806 }{ 0.873 }{ 93.2\% }{ 100.0\% }
\\

\end{tabular}
\end{center}
    \caption{Terrains generated with the CA rules described in Table~\ref{tab:rules} after $i$ iterations. The initial state, generated with random noise ($seed=123$) at $i=0$, is discarded. All heightmaps are normalized to $[0, h_{\max}]$ via min-max scaling, where $h_{\max}$ is given as a percentage of the heightmap resolution, $129 \times 129$---which also corresponds to the CA dimensions.
    The following metrics are present below each terrain: roughness index ($R$), normalized entropy ($E$), percentage of walkable areas ($W$), and path coverage ($W_0$).
    }
    \label{fig:experiments}

\end{figure*}

\section{Results, Discussion, and Limitations}
\label{sec:results}

Fig.~\ref{fig:experiments} shows terrains generated with Algorithm~\ref{alg:hmgen} using various CA rules after the specified iterations. Majority rules produce contrasting terrains depending on parameters. Setting the Majority radius to 2, as in the first row of Fig.~\ref{fig:experiments}, produces interesting results. For $i=5$, the terrain resembles an eroded landscape with abrupt cliffs and mesas; at $i=20$, it becomes more diverse, mostly hills with some sharp cliffs. The number of iterations offers a consistent parameter for controlling terrain smoothness. This is reflected by a steady roughness index decrease (0.011 at $i=3$ to 0.003 at $i=50$), indicating smoothing. Concurrently, normalized entropy increases (0.211 to 0.366), reflecting greater elevation diversity. Walkability improves markedly (29.4\% to 90.5\%), and path coverage reaches 100\%, showing improved navigability as the terrain evolves.

Continuing with the Majority rule, setting $r=4$ and $N=38$ yields a landscape with irregular holes, potentially useful as a surface for worn-out objects. Comparing it to the previous case of $r=2$, it is possible to conclude that in this case the main terrain features stand out sooner (i.e., for lower $i$), while maintaining their prominence and various minor irregularities for longer (i.e., for higher $i$). Quantitatively, these terrains display a similarly declining roughness index, from 0.017 at $i = 3$ to 0.005 at $i = 20$, with entropy values increasing modestly (from 0.191 to 0.270). This suggests a terrain that is less topographically varied but becomes smoother over time. Walkability and path coverage increase significantly after $i = 10$, peaking at 78.4\% and 97.2\%, respectively, indicating a threshold after which terrains become highly accessible.

Results for the Caves rule are particularly interesting, as they closely follow their 2D counterpart. The rule was tested with $r=2$ and neighbor threshold $N=13$, with results shown in the third row of Fig.~\ref{fig:experiments}. From a stacked 3D perspective, the 2D caves become eroded landscapes, increasingly well defined during the iterative process. Contrary to the similarly parameterized Majority rule ($r=2$, $N=13$), the eroded landscape is cleaner and holds its shape for longer. From a metrics standpoint, the roughness remains stable around 0.007--0.010, while entropy increases from 0.214 to 0.434, revealing growing structural complexity. Walkability also improves significantly (32.0\% to 80.0\%), and path coverage rises sharply, peaking at 94.0\% at $i = 20$. These values support the conclusion that Caves rule yields terrain that is both expressive and increasingly navigable over time.

Finally, the chaotic Diamoeba rule is also able to produce natural-looking landscapes. While terrains seem somewhat rugged for $i<50$, the rule performs best for $i \geq 100$, yielding diverse surfaces with hills, canyons, mesas, as well as smaller yet smooth features. 
Objectively, this evolution is clearly reflected in the metrics: roughness drops from 0.009 to 0.002 between $i = 5$ and $i = 200$, while entropy grows from 0.309 to a high of 0.873, showing strong diversification in terrain structure. Walkability increases from 5.2\% to 93.2\%, and path coverage reaches full connectivity at 100\%, confirming that Diamoeba generates highly detailed yet playable terrains with sufficient iterations.

\begin{figure*}[t]
\begin{center}
\begin{tabular}{ccccc}
\includegraphics[width=3.25cm]{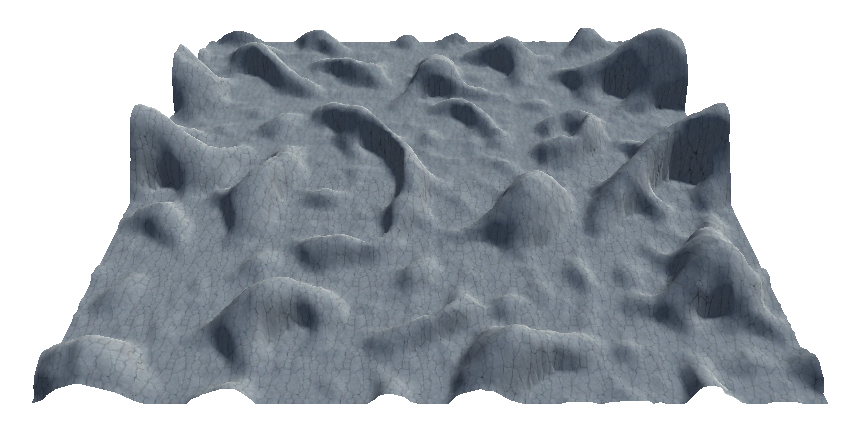}  &
\includegraphics[width=3.25cm]{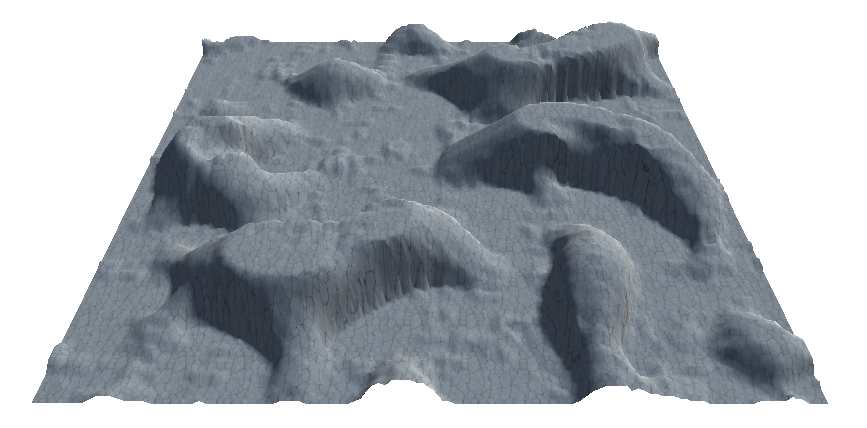}  &
\includegraphics[width=3.25cm]{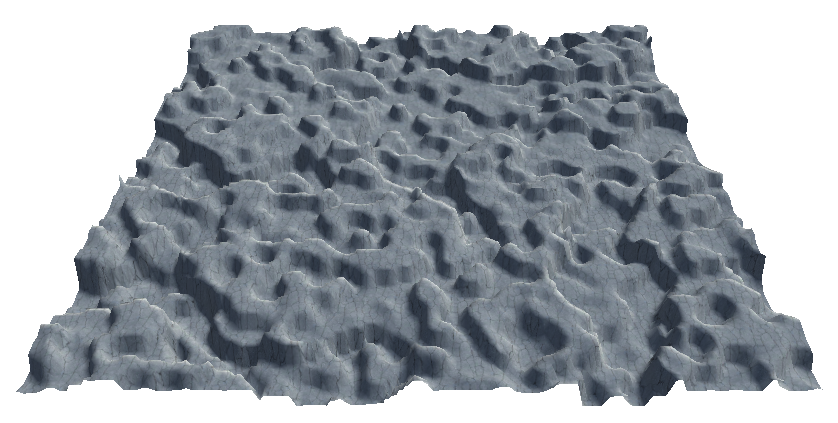}  &
\includegraphics[width=3.25cm]{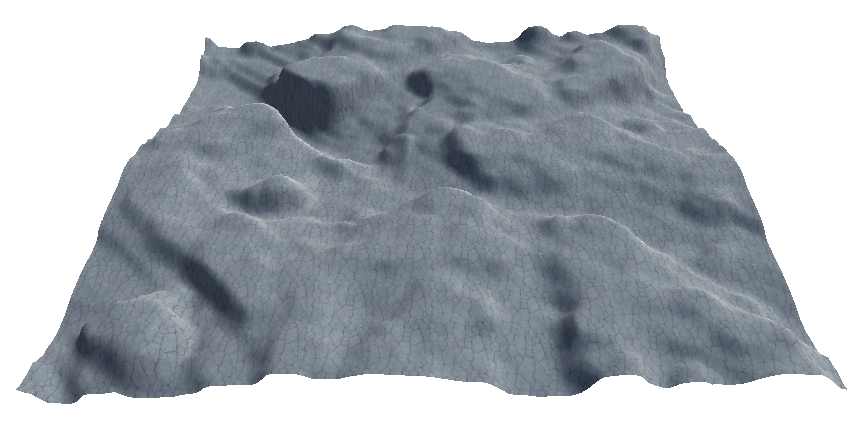}  &
\includegraphics[width=3.25cm]{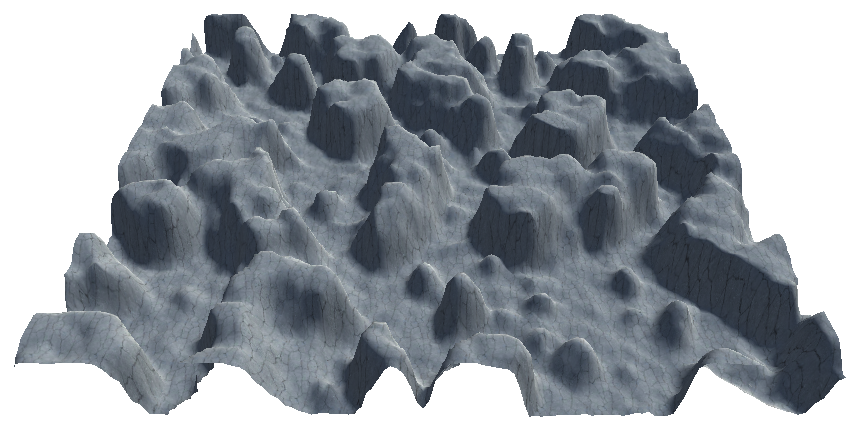}  \\
\metic{ 0.006 }{ 1.725 }{ 0.414 }{ 68.4\% }{ 99.0\% }
&
\metic{ 0.007 }{ 1.638 }{ 0.368 }{ 72.6\% }{ 99.8\% }
&
\metic{ 0.007 }{ 1.774 }{ 0.686 }{ 45.3\% }{ 11.3\% }
&
\metic{ 0.003 }{ 1.807 }{ 0.887 }{ 83.7\% }{ 100.0\% }
&
\metic{ 0.009 }{ 1.801 }{ 0.639 }{ 41.1\% }{ 36.6\% }
\\
{\footnotesize(a)} & {\footnotesize(b)} & {\footnotesize(c)} & {\footnotesize(d)} & {\footnotesize(e)}
\end{tabular}
\end{center}
     \caption{Additional experiments highlighting the potential of the proposed technique:
  (a) Majority, $i=50$, $r=2$, $N=13$, $h_{\max}=12.5\%$, $seed=700$;
  (b) Majority, $i=20$, $r=4$, $N=41$, $h_{\max}=10\%$, $seed=123$;
  (c) Three layers of Caves added together:
    i)  $i=3$, $r=1$, $N=5$, $h_{\max}=0.5\%$, $seed=123$; 
    ii) $i=50$, $r=1$, $N=5$, $h_{\max}=1.5\%$, $seed=123$; 
    and,
    iii) $i=50$, $r=2$, $N=13$, $h_{\max}=3\%$, $seed=500$; 
  (d) Diamoeba, $i=500$, $r=1$, $h_{\max}=10\%$, $seed=123$; and, (e) Diamoeba, $i=50$, $r=1$, $h_{\max}=10\%$, $seed=700$.
   The following metrics are present below each terrain: roughness index ($R$), normalized entropy ($E$), percentage of walkable areas ($W$), and path coverage ($W_0$).
  }
  \label{fig:other}

\end{figure*}

Fig.~\ref{fig:other} shows additional results one can obtain with the technique. Fig.~\ref{fig:other}(a) displays a Majority rule similar to what is shown in the top row, rightmost column of Fig.~\ref{fig:experiments} ($r=2$, $N=13$, $i=50$). The difference is in the seed used for generating the initial noise and the larger $h_{\max}$, resulting in a generally soft but distinct alien-looking landscape with prominent features. This example exhibits low roughness (0.006), moderate entropy (0.414), and high walkability (68.4\%) and coverage (99.0\%), making it a well-connected, explorable terrain. Fig.~\ref{fig:other}(b) highlights a Majority rule with $r=4$ and $N=41$, 
producing a mix of mostly flat terrain with extrusive and well-defined features. Compared with the previous example, this parameterization yields slightly higher roughness (0.007) but strong metrics overall (72.6\% walkability and 99.8\% coverage), suggesting flat but structured topography.

The landscape shown in Fig.~\ref{fig:other}(c) depicts a combination of three separate terrains added together. These were generated with different parameterizations of the Caves rule. The first two layers, defining the finer details of the landscape, used $r=1$, $N=5$, and $seed=123$; the first one is obtained from $i=3$ with a very small $h_{\max}$, while the second was collected with $i=50$ and scaled to a larger height. The third layer, generated with $r=2$, $N=5$, a different seed, and higher $h_{\max}$ than the previous two, sets up the coarser aspects of the landscape. Although this combination displays high entropy (0.686), it has relatively low walkability (45.3\%) and very limited path coverage (11.3\%), suggesting fragmented topographical complexity.

Finally, Fig.~\ref{fig:other}(d) and Fig.~\ref{fig:other}(e) offer two additional perspectives on the Diamoeba rule: the former is a steeper, further iterated version of the natural-looking Diamoeba landscape already presented in Fig.~\ref{fig:experiments} (bottom row, rightmost image), while the latter also increases height but uses a different seed to produce a rocky-like terrain with various features. In terms of metrics, Fig.~\ref{fig:other}(d) is highly optimized ($R = 0.003$, $E = 0.887$), showing excellent walkability (83.7\%) and perfect connectivity (100\%), while Fig.~\ref{fig:other}(e) displays more rugged terrain ($R = 0.009$), with lower walkability (41.1\%) and limited connectivity (36.6\%), emphasizing the importance of the number of stacked CA iterations.

We believe the preliminary results presented here are interesting in themselves, some with aesthetically pleasing features, others doing an arguably good job of mimicking real world landscapes or surfaces, and many displaying high levels of objective diversity combined with large playable areas from a pathfinding standpoint. However, these results barely scratch the surface of what is possible with the proposed technique. Among innumerable CA transition rules, only a few were experimented with here, and all of them seeded with an initial grid of random noise. Tweaking the initial probability of live cells will surely yield distinct results, as well as using predefined initial shapes, which some rules respond better to in the 2D case~\cite{toffoli1987cellular}. As shown in the example of Fig.~\ref{fig:other}(c), combining together different heightmap generators 
holds the potential for further customization of the produced landscapes. Finally, 2D CAs with small neighborhoods offer good performance and can be GPU-parallelized \cite{cagigas2022efficient}, making them suitable for real-time map and level generation~\cite{johnson2010cellular}.

This work presents some limitations. Although it includes quantitative terrain analysis with established metrics, no comparison is made with other techniques, as this paper mainly demonstrates the viability of the proposed method. Such comparisons are essential to assess competitiveness and generalizability. Due to their generative nature, CAs suit \textit{generate-and-test} scenarios, where terrains are iteratively produced and evaluated against thresholds for metrics such as roughness or walkability~\cite{yannakakis2018artificial}. Alternatively, CA parameters (e.g., radius, thresholds, $i$, $h_{\text{max}}$) could be optimized via search techniques (e.g., genetic algorithms) to target specific metrics~\cite{yannakakis2018artificial}. Finally, while we observed how metrics evolve with iteration count, the effects of other parameters on aesthetics and quality remain largely unexplored. Understanding these dependencies is crucial to improve control and usability in practical applications.

\section{Conclusions and Future Work}
\label{sec:conclusions}

We presented a simple yet effective method for generating 3D terrains by accumulating iterations of 2D cellular automata into a heightmap. The approach yields diverse and visually compelling results, also supported by quantitative metrics capturing roughness, entropy, walkability, and connectivity. Future work includes systematic exploration of CA rules and parameters, combining different generators, and comparing this method against established terrain generation techniques. Optimization-based control over terrain features, as well as real-time applications, are also promising directions.

\bibliographystyle{IEEEtran}

\end{document}